\documentstyle[12pt]{article}
\begin{document}
\baselineskip18pt
\title{Noise
perturbations in the Brownian motion  and quantum dynamics}

\author{Piotr Garbaczewski\\
 Institute of Physics, Pedagogical University,\\
  pl. S{\l}owia\'{n}ski 6, PL-65 069 Zielona G\'{o}ra,
  Poland\thanks{Email:  pgar@omega.im.wsp.zgora.pl}}
\maketitle
\begin{abstract}
The third Newton law for  mean velocity fields is utilised
to generate anomalous (enhanced) or  non-dispersive
diffusion-type processes which, in particular,  can be
interpreted as a probabilistic counterpart of the
Schr\"{o}dinger picture  quantum dynamics.
\end{abstract}
\vskip2cm

If we consider  a   fluid in thermal equilibrium as the noise
 carrier, a kinetic theory viewpoint  amounts to visualizing the
 constituent molecules that collide
not only with each other but  also  with the tagged (colloidal)
particle, so \it enforcing \rm  and \it maintaining  \rm
its  observed erratic  motion.
The Smoluchowski  approximation  takes  us   away
 from those kinetic theory intuitions by  projecting  the
phase-space theory of random motions into its  configuration
space image  which is a spatial Markovian diffusion
process, whose formal infinitesimal encoding reads:
$${d\vec{X}(t)= {\vec{F}\over {m\beta }}dt +
\sqrt{2D}d\vec{W}(t)\enspace .}\eqno (1)$$

In the above $m$ stands for the mass of a diffusing particle,
$\beta $ is a friction parameter, D is a diffusion constant
 and $\vec{W}(t)$ is a normalised Wiener process.
The Smoluchowski forward drift can be traced back to
a presumed selective action of the external force $\vec{F}=
-\vec{\nabla }V$ on the Brownian particle
that  has a negligible  effect on the thermal bath but
in view of frictional  resistance imparts to a particle
the \it  mean \rm velocity $\vec{F}/m\beta $ on
the $\beta ^{-1}$ time scale, \cite{nel}.  The
noise carrier (fluid in the present considerations)
statistically remains in the state of rest, with \it no \rm
intrinsic mean flows.

An implicit phase-space scenario  of the Brownian motion
 refers  to  minute acceleration/deceleration events  which
modify (infinitely  gently, \cite{nel}, at a generic
rate of $10^{21}$ times
 per second) velocities of  realistic    particles.
Clearly, the microscopic
energy-momentum  conservation laws need to be respected in
each separate collision event.  In contrast to derivations based
on the Boltzmann equation, this feature is completely
\it  alien \rm to the Brownian motion theory.
That energy-momentum deficit is one
of "forgotten" or tacitly "overlooked"  problems in the
standard  theory of the Brownian motion, see
e.g. \cite{vigier,pre99}.

An issue of \it origins \rm  of general isothermal flows in the
noisy environment is far from being settled, \cite{pre99}.
Namely, the standard theory  does not account
(or refers to phenomena for  which that is not necessary)
for  generic  perturbations  of the random
medium  as a reaction to the  enforced (exclusively due to the
impacts coming from the medium) particle motion.
An ability of the medium to perform  work  i. e.
to give  a kinetic energy to Brownian particles (strictly speaking,
in the local mean, hence on the ensemble average) appears to be
a universal feature of the thermal bath even in the absence
of any external force.

Then, there arise problems  with the thermal equilibrium,
\cite{pre99}.
The tagged particle always propagates "at the expense" of the bath,
which nonetheless  remains "close" to its thermal equilibrium.
In view of the involved local cooling and heating phanomena,
which are  necessary to keep in conformity  with the second
law of thermodynamics, the bath should develop local flows and thus
 in turn actively \it react back  \rm to what is being
happenning to the particle in the course of its   random propagation.
Such environmental reaction, while interpreted as a feedback effect,
surely can be neglected in   an individual particle propagation
sample.
However, an obvious  Brownian particle energy-momentum
"deficit" on the ensemble  average (kinetic energy  and momentum
is carried locally by diffusion currents), indicates that
the  medium  reaction
may give observable contributions on the ensemble average,
as a  statistically accumulated feature.
For a statistical ensemble of weakly-out-of-equilibrium systems,
an effective isothermal scenario can be used again, but then
mean flows in the noise carrier are "born" as a consequence of
the averaging, \cite{pre99}.

It is well known that
a spatial diffusion (Smoluchowski) approximation of the
phase-space process, allows
to reduce the number of \it independent \rm
local conservation laws
(cf. \cite{zambrini}-\cite{guth}) to two only.
Therefore the Fokker-Planck  equation can
always be supplemented  by another (independent) partial
differential equation to form a \it closed \rm system.

 If we assign
 a probability density $\rho _0(\vec{x})$ with which the
 initial data
  $\vec{x}_0=\vec{X}(0)$ for Eq. (1) are
distributed, then the emergent Fick law  would  reveal a
statistical tendency of particles to flow away from
higher probability  residence areas.     This
feature is encoded in the corresponding Fokker-Planck equation:
$${\partial _t \rho  = - \vec{\nabla }\cdot (\vec{v}\rho )=
- \vec{\nabla }\cdot [({\vec{F}\over {m\beta }}  - D
{{\vec{\nabla }\rho }\over {\rho }}) \rho ]}\eqno (2)$$
where a diffusion current velocity  is
$\vec{v}(\vec{x},t) = \vec {b}(\vec{x},t) - D{{\vec{\nabla }\rho
(\vec{x},t)}\over {\rho (\vec{x},t)}}$
while  the forward drift reads $\vec{b}(\vec{x},t) =
{\vec{F}\over {m\beta }}$.
Clearly, the local diffusion current (a local flow that might
be experimentally observed for  a cloud of suspended particles
in a liquid)
$\vec{j}=\vec{v} \rho $
gives rise to a non-negligible  matter transport  on the
ensemble average, \cite{blanch,pre99}.

It is interesting to notice that the local velocity field
 $\vec{v}(\vec{x},t)$
  obeys the natural (local) momentum  conservation law
which   directly originates from
the rules of the It\^{o} calculus for Markovian diffusion processes,
\cite{nel}, and from the first moment equation in the
diffusion approximation  (!) of the Kramers theory, \cite{blanch}:
$${\partial _t\vec{v} + (\vec{v} \cdot \vec{\nabla }) \vec{v} =
\vec{\nabla }(\Omega - Q)\enspace .}\eqno (3)$$

An effective  potential function $\Omega (\vec{x})$
can be expressed in terms of the  forward drift
$\vec{b}(\vec{x}) = {\vec{F}(\vec{x})
\over {m\beta }}$ as follows:
$\Omega = {{\vec{F}^2} \over {2m^2\beta ^2}} + {D\over {m\beta }}
\vec{\nabla } \cdot \vec{F}$.

Let us emphasize that it is the diffusion (Smoluchowski)
approximation which makes
the right-hand-side of Eq. (3) substantially  different from the
usual moment equations appropriate for  the Brownian motion.
In particular, the force $\vec{F}$ presumed to act
upon an individual particle, does not give rise in Eq. (3)
to the  expression
$-{1\over m}\vec{\nabla }V$  which might be expected on the basis of
kinetic theory intuitions and moment identities directly derivable
from the Karmers equation, but to  the term
$+\vec{\nabla }\Omega $.

Moreover, instead of the standard pressure term,
 there appears a contribution from a
  probability density  $\rho $-dependent potential
 $Q(\vec{x},t)$. It  is given in terms of the so-called osmotic
velocity field  $\vec{u}(\vec{x},t) =
D\vec{\nabla } \, ln \rho (\vec{x},t)$, (cf. \cite{nel}):
$Q(\vec{x},t) = {1\over 2} \vec{u}^2 + D\vec{\nabla }
\cdot \vec{u}$
and  is  generic to a local momentum conservation
law  respected by   isothermal Markovian diffusion processes, cf.
 \cite{nel}-\cite{guth}.

The Smoluchowski drift \it does not \rm refer to any  flows \it
in \rm the noise carrier.
To  analyze   perturbations  of the
medium and then the resulting intrinsic (mean) flows,
a  more general function
$\vec{b}(\vec{X}(t),t)$,  must  replace the Smoluchowski drift in
Eqs. (1), (2).
Forward drifts modify additively the pure noise
(Wiener process entry) term in the It\^{o} equations.
Under suitable restrictions, we can relate probability  measures
corresponding to  different (in terms of forward drifts !)
Fokker-Planck equations  and  processes by means of the
Cameron-Martin-Girsanov theory  of measure transformations.
The Radon-Nikodym derivative of measures is here involved and
for suitable  forward drifts which are
are gradient fields   that  yields, \cite{blanch},
the most  general  form of an auxiliary potential
$\Omega (\vec{x},t)$ in Eq. (3):
$${\Omega (\vec{x},t) = 2D[ \partial _t\phi + {1\over 2}
({\vec{b}^2\over {2D}} + \vec{\nabla }\cdot \vec{b})]\enspace .}
\eqno (4)$$
We denote $\vec{b}(\vec{x},t) = 2D \vec{\nabla } \phi (\vec{x},t)$.

Eq.  (4) is a   trivial identity,
if we take for granted that
all drifts are known from the beginning, like in case of typical
Smoluchowski diffusions where the external force $\vec{F}$ is
a priori postulated.
We can proceed otherwise and,
 on the contrary, one  can  depart
from a  suitably chosen    space-time dependent function
$\Omega (\vec{x},t)$.
From this point of view, while developing the formalism, one should
decide what is a quantity of a \it primary \rm physical interest:
the field  of drifts $\vec{b}(\vec{x},t)$ or the potential
$\Omega (\vec{x},t)$.

   Mathematical features of the formalism
appear to depend crucially on the properties (like continuity,
local and global boundedness, Rellich class) of the potential
$\Omega $, see e.g. \cite{blanch,carlen}.
Let us consider
a bounded from below (local boundedness from above is useful as well),
 continuous function $\Omega (\vec{x},t)$,
cf. \cite{blanch}).  Then,  by means of the  gradient
field ansatz for the diffusion current velocity
($\vec{v}=\vec{\nabla }S\rightarrow
\partial _t\rho = - \vec{\nabla }\cdot [(\vec{\nabla } S)\rho ]$)
we can  transform the momentum conservation law (3) of a
Markovian diffusion process to  the universal
Hamilton-Jacobi form:
$${\Omega = \partial _tS + {1\over 2} |\vec{\nabla }S|^2  + Q }
\eqno (5)$$
where $Q(\vec{x},t)$ was defined before. By
applying the  gradient operation to Eq. (5) we recover (3).
In the above,  the contribution due to $Q$
is a direct consequence of  an initial probability measure choice
for the diffusion    process,
 while $\Omega $ via Eq. (4)  does account for an appropriate
forward drift of the process.

Thus, in the context of Markovian diffusion processes,
 we can consider a  closed system of partial differential equations
 which comprises   the continuity
equation $\partial _t \rho =- \vec{\nabla }(\vec{v}\rho )$
 and  the Hamilton-Jacobi  equation
(5), plus suitable initial (and/or boundary) data.
The underlying diffusion process is specified uniquely, cf.
\cite{blanch,zambrini}.

Since the pertinent nonlinearly coupled system  equations looks
discouraging,
it is useful to mention  that a  \it
linearisation \rm of this
 problem is provided by a time-adjoint pair of
generalised diffusion equations  in the framework
 of the so-called Schr\"{o}dinger boundary data problem,
 \cite{blanch,zambrini}.
  The standard heat equation appears as a very special
  case in this formalism.

 The local conservation law  (3)  acquires a direct
 physical meaning (the
rate of change of momentum
carried by a locally co-moving with the flow volume, \cite{blanch}),
only if averaged with respect to $\rho (\vec{x},t)$ over
a   simply connected spatial area.   If $V$ stands for
a volume enclosed by a two-dimensional outward oriented
surface $\partial V$, we define a co-moving
volume on small time scales, by deforming the boundary surface in
accordance with the local current velocity  field values.
Then , let us consider  at time $t$ the  displacement of the
boundary surface $\partial V(t)$ defined  as follows:
$\vec{x}\in \partial V \rightarrow \vec{x} +
\vec{v}(\vec{x},t)\triangle t$ for all $\vec{x}\in \partial V$.
Up to the first order in $\triangle t$  this guarantees the
conservation of mass (probability measure)  contained
in $V$ at time $t$ i. e. $\int_{V(t+\triangle t)}\rho
(\vec{x},t+\triangle t)d^3x  - \int_{V(t)}\rho (\vec{x},t)d^3x
\sim 0$.

The  corresponding (to the leading order in $\triangle t$)
quantitative  momentum   rate-of-change measure reads,
cf. \cite{blanch},
$\int_V \rho \vec{\nabla }(\Omega - Q)d^3x$.

 For a particular  case of the
 free  Brownian expansion  of an initially given
 $\rho _0(\vec{x})=
{1\over {(\pi \alpha ^2)^{3/2}}} exp
(-{x^2\over \alpha ^2}) $,  where $\alpha ^2=4Dt_0$,
 we would have
 $ -\int_V\rho \vec{\nabla }Q d^3x=-
\int_{\partial V}Pd\vec{\sigma }$, where
$Q(\vec{x},t) =
{\vec{x}^2\over {8(t+t_0)^2}} - {{3D}\over {2(t+t_0)}}$, while
the "osmotic pressure" contribution reads  $P(\vec{x},t)= -{{D}
\over {2(t+t_0)}} \rho (\vec{x},t)$ for all
$\vec{x}\in R^3$ and $t \geq 0$.

The current velocity   $\vec{v}(\vec{x},t)=\vec{\nabla }
S(\vec{x},t)=
{\vec{x}\over {2(t+t_0)}}$ is  linked to the Hamilton-Jacobi
equation
$\partial _tS +{1\over 2}|\vec{\nabla }S|^2 + Q =0$
whose solution reads: $S(\vec{x},t)= {\vec{x}^2\over {4(t+t_0)}}
 + {3\over 2}D ln[4\pi D(t+t_0)]$.

Let us observe that the initial data $\vec{v}_0=-D\vec{\nabla }
ln\, \rho _0= -\vec{u}_0$ for the current
velocity field  indicate that we have totally  ignored
a crucial  \it preliminary \rm stage of the dynamics on the
$\beta ^{-1}$ time scale,  when the Brownian expansion of an
initially \it static \rm  ensemble has been ignited and so particles
have  been ultimately  set  in motion.

Notice also that our "osmotic expansion pressure" $P(\vec{x},t)$
is not positive definite, in contrast to
 the familiar kinetic theory (equation of state) expression
 for the pressure  $P(\vec{x})= \alpha \,
 \rho ^{\beta }(\vec{x}), \,  \alpha >0$ appropriate for gases.
 The admissibility of the negative sign of the "pressure"  function
 encodes the fact that the
Brownian evolving concentration of particles generically
decompresses (blows up),
instead of being compressed by the surrounding medium.

The loss  (in view of the "osmotic" migration) of momentum
 stored in a  control volume at a given time,
may be here interpreted
in terms of  an acceleration   $-\int_V\rho \vec{\nabla }Qd^3x$
induced by a \it fictituous \rm "attractive force".

By invoking an  explicit Hamilton-Jacobi connection (5),
we may attribute
to a diffusing Brownian  ensemble
 the mean kinetic energy   per unit of mass
$\int_V\rho {1\over 2}\vec{v}^2 d^3x$.
In view of $<\vec{x}^2>=6D(t+t_0)$,  we have also
$\int_{R^3}\rho {1\over 2}\vec{v}^2 d^3x= {{3D}\over {4(t+t_0)}}$.
Notice that the mean energy
$\int_V\rho ({1\over 2}\vec{v}^2+Q) d^3x$  needs not to be  positive.
Indeed, this expression identically vanishes after extending
integrations from $V$ to $R^3$.
On the other hand the kinetic contribution, initially equal
$\int_{R^3}{1\over 2} \rho v^2 d^3x = 3D/\alpha ^2$ and
evidently \it coming from nowhere, \rm continually
diminishes and is bound to disappear in the asymptotic
$t\rightarrow \infty $ limit, when  Brownian particles become
uniformly distributed in space.

Normally, diffusion processes yielding a nontrivial matter
transport (diffusion currents) are observed
for a  non-uniform  concentration of colloidal particles which
are regarded as independent (non-interacting).
We can however devise a thought (numerical) experiment
that gives rise
to a corresponding transport in terms of an ensemble of
sample (and thus independent) Brownian motion realisations on a
fixed finite  time interval,
 instead of considering a multitude of them (migrating swarm of
 Brownian particles)  simultaneously.

 Let us assume that
"an effort" (hence, an energy loss/gain  which implies local
deviations from thermal equilibrium conditions) of the random medium,
on the
$\beta ^{-1}$ scale, to produce a local Brownian diffusion current
$(\rho \vec{v})(\vec{x},t_0)$  out of the  initially static ensemble
and thus to decompress (lower the blow-up tendency) an initial
non-uniform probability distribution, results in the \it effective
osmotic   reaction \rm of  the random medium.
   This  is the \it Brownian recoil effect \rm of Ref. \cite{vigier}.

In that case, the particle swarm   propagation scenario becomes
entirely different from the standard  Brownian one.
First of all, the nonvanishing forward drift   $\vec{b}=\vec{u}$
is  generated as a dynamical (effective, statistical here !)
response of the bath to the enforced by the bath  particle
transport with the local
velocity $\vec{v}= -\vec{u}$.
Second, we need to account for a parellel  inversion of
the pressure effects (compression $+\vec{\nabla }Q$ should
replace the decompression $-\vec{\nabla }Q$) in the respective local
momentum conservation law.

Those features can be secured    through an explicit  realization
 of  the action-reaction principle
 which we promote to the status of the \it  third Newton law
 in the mean. \rm

On the level of Eq. (3), once averaged over a finite
volume, we interpret the momentum per unit of mass rate-of-change
$\int_V \rho \vec{\nabla }(\Omega - Q)d^3x$   which occurs
exclusively due to the Brownian expansion mechanism, to generate
a counterbalancing rate-of-change tendency in the random medium.
To account for the emerging forward drift and an obvious
modification  of the subsequent expansion  of an ensemble of
 particles, we  re-define Eq. (3) by setting
  $- \int_V \rho \vec{\nabla }(\Omega - Q)d^3x$ in its
  right-hand-side instead of  $+ \int_V \rho \vec{\nabla }
  (\Omega - Q)d^3x$ .   That amounts to   an
  instantaneous  implementation  of
  the third Newton law in the mean (action-reaction principle)
  in Eq. (3).

Hence, the momentum conservation law for the process \it
with a recoil \rm
(where the reaction term replaces the decompressive
"action" term) would read:
$${\partial _t\vec{v} + (\vec{v}\cdot \vec{\nabla })\vec{v} =
\vec{\nabla } (Q- \Omega )}\eqno (6) $$
so  that
$${\partial _t S + {1\over 2}
|\vec{\nabla }S|^2 - Q= -\Omega }\eqno (7)$$
stands for the corresponding Hamilton-Jacobi equation, cf.
\cite{zambrini,hol}, instead of Eq. (5).
A suitable adjustment (re-setting) of the initial data is
here necessary.

In the coarse-grained picture of motion we thus  deal with a
sequence    of repeatable scenarios realised
on the  Smoluchowski process  time scale $\triangle t$:
the Brownian swarm expansion build-up is accompanied by the parallel
counterflow build-up, which in turn modifies the subsequent
stage of the Brownian swarm migration  (being interpreted to modify
the forward drift of the process) and the corresponding
built-up anew counterflow.

The new closed system of partial differential equations
refers to Markovian diffusion-type  processes
again, \cite{nel,blanch,carlen}.
The link is particularly obvious if we observe that the
 new Hamilton-Jacobi equation (7) can be formally
rewritten in the previous form (5) by introducing:
$${\Omega _r= \partial _tS + {1\over 2} |\vec{\nabla }S|^2 + Q }
\eqno (8)$$
where $\Omega _r= 2Q - \Omega $ and
 $\Omega $ represents the previously defined potential function
of any Smoluchowski (or more general) diffusion process.

It  is $\Omega _r$  which via Eq. (4) would determine forward drifts
of the Markovian diffusion process with a recoil. They must obey the
Cameron-Martin-Girsanov identity
$\Omega _r = 2Q- \Omega =
2D[ \partial _t\phi + {1\over 2}
({\vec{b}^2\over {2D}} + \vec{\nabla }\cdot \vec{b})]$.

Our new closed system of equations is badly nonlinear and coupled,
but its linearisation
can be immediately given in terms of an adjoint pair of
Schr\"{o}dinger equations with  a potential $\Omega $,
\cite{nel,zambrini}.
Indeed,
$i\partial _t \psi = - D\triangle \psi + {\Omega \over {2D}}\psi $
with a solution $\psi = \rho ^{1/2} exp(iS)$
and its complex adjoint makes the job, if
we regard  $\rho $ together with $S$ to remain in conformity with
the previous notations. The choice of
$\psi (\vec{x},0)$ gives rise to a solvable Cauchy problem.
(Notice that by setting $D={\hbar \over {2m}}$ we recover 
the standard quantum mechanical notation.) 

This feature we shall exploit in below.
Notice that, for time-indepedent $\Omega $, the
total energy $\int_{R^3}({v^2\over 2} -Q + \Omega )\rho d^3x$
 of the diffusing ensemble is a  conserved quantity.

  The general existence criterions for Markovian
diffusion   processes of that kind, were formulated in Ref.
\cite{carlen}, see also \cite{zambrini,blanch}.

Let us  consider a  simple one-dimensional example.
In the absence of external forces, we solve the
equations   (in  space dimension one)  $\partial _t\rho =
-\nabla (v\rho )$ and $\partial _t v + (v\nabla )v = + \nabla Q$,
with an initial probability density $\rho _0(x)$  chosen in
correspondence with the previous free Brownian motion example.
We denote $\alpha ^2= 4Dt_0$.
Then,
$\rho (x,t)=
{\alpha \over {[\pi (\alpha ^4 + 4D^2t^2)]^{1/2}}}\,
exp[-{{x^2\alpha ^2}\over  {\alpha ^4 + D^2t^2}}]$
and
$b(x,t)= v(x,t) + u(x,t)= {{2D(\alpha ^2 - 2Dt)x} \over
{\alpha ^4 +  4D^2t^2}}$
are the pertinent solutions.
Notice that $u(x,0)= -{{2Dx}\over \alpha ^2}=b(x,0)$  amounts to
$v(x,0)=0$,  while in the previous free Brownian case the initial
current velocity was equal to $-D\nabla ln\, \rho _0$.
This re-adjustment of the initial data can be interpreted in
terms of the counterbalancing (recoil) phenomenon:
the  would-be initial Brownian
ensemble current velocity  $v_0=-u_0$ is here
completely saturated by the emerging forward  drift
$b_0=u_0$, see e.g. also \cite{vigier}.
We deal also  with a fictituous "repulsive" force,
which corresponds to the compression (pressure upon) of
the Brownian ensemble due to the counter-reaction of the
surrounding  medium.
We can write things more explicitly. Namely, now:
$Q(x,t)= {{2D^2\alpha ^2}\over {\alpha ^4 + 4D^2t^2}}
({{\alpha ^2 x^2}\over {\alpha ^4+ 4D^2t^2}} - 1)$
and the corresponding pressure term ($\nabla Q=
{1\over \rho }\nabla P$) reads
$P(x,t) = - {{2D^2\alpha ^2}\over {\alpha ^4 + 4D^2t^2}}
\rho (x,t)$
giving a positive contribution $+\nabla Q$ to the local
conservation law.
The    related Hamilton-Jacobi equation
$\partial _tS + {1\over 2} |\nabla S|^2 = + Q$
is solved by
$S(x,t) = {{2D^2x^2t}\over {\alpha ^4 + 4D^2t^2}} - D\,
arctan\, (-{{2Dt}\over \alpha ^2})$.
With  the above form of $Q(x,t)$ one can readily check that
 the Cameron-Martin-Girsanov constraint euqation for the
forward drift of the Markovian diffusion process with a recoil
is automatically valid for $\phi  ={1\over 2}ln\, \rho  + S$:
$2Q = 2D[\partial _t\phi +{1\over 2}({b^2 \over {2D}} +
\nabla \cdot b)]$.

In anology with our free Brownian motion discussion, let us observe
that  presently
$<x^2> = {\alpha ^2\over 2} + {{2D^2t^2}\over \alpha ^2}$.
It is easy to demonstrate that the quadratic dependence on time
persists for arbitrarily  shaped  initial choices of the
probability distribution $\rho _0(x)>0$.
That signalizes an  anomalous behaviour (enhanced diffusion)
of the pertinent Markovian  process when $\Omega =0$
i. e. $\Omega _r=2Q$.

We can evaluate the kinetic energy contribution
$\int_{R} \rho {v^2\over 2}dx = {{4D^4t^2}
\over {\alpha ^2(\alpha ^4 + 4D^2t^2)}}$
which in contrast to the Brownian case shows up a continual growth
up to the terminal (asymptotic) value ${D^2\over \alpha ^2}$.
This value was in turn an initial kinetic  contribution
in  the previous free Brownian expansion example.
In contrast to that case, the total energy integral is now finite
(finite energy diffusions of Ref. \cite{carlen}) and reads
$\int_R({1\over 2}v^2 - Q)\rho dx = {D^2\over \alpha ^2}$
 (it  is a conservation law).
The asymptotic value of the current velocity
$v\sim {x\over t}$  is twice larger than   this appropriate for the
Brownian motion, $v\sim {x\over {2t}}$.

 It is easy to
 produce an  explicit solution  to  (7), (8)
 in case of $\Omega (x)=
{1\over 2}\gamma ^2x^2   - D\gamma $,
with  exactly the same inital probability density
$\rho _0(x)$ as before.
The forward drift of the corresponding diffusion-type process
does not show up  any obvious  contribution from the
harmonic Smoluchowski force. It
  is  completely eliminated by the Brownian recoil scenario.
One may check that  $b(x,0)= -{{2Dx}\over \alpha ^2} = u(x,0)$,
while  obviously $b={F\over {m\beta }}= -\gamma x$ would hold
true for all times, in case of the Smoluchowski
diffusion process.
Because  of the harmonic attraction  and suitable initial
probability measure choice, we have here
wiped out all  previously  discussed enhanced diffusion
features.
Now,  the dispersion  is attentuated and actually  the
non-dispersive diffusion-type  process is  realised:
$<x^2>$ does not spread at all despite of the intrinsically
stochastic  nature of the dynamics (finite-energy diffusions
of Ref. \cite{carlen}).

It is clear that stationary processes \it are the same \rm
both in case of  the  standard Brownian motion  and the
Brownian motion with  a recoil. The respective
propagation scenarios substantially  differ in the
non-stationary case only.

One may possibly argue, that a sign  inversion   of the
right-hand-side of Eq. (3) which implies (6), can be accomplished
by means of an analytic continuation in time (cf. the Euclidean
quantum  mechanics discussion in Ref. \cite{zambrini}). An examination
of free dynamics cases  given in the above proves that the third
Newton law ansatz is an independent procedure.

{\bf Acknowledgements:}
I would like to thank Professor Eric Carlen for discussion
about the scaling limits of the Boltzmann equation and  related
conservation laws.


\begin{thebibliography}{99}


\bibitem{nel}   E. Nelson, "Dynamical Theories of the Brownian
Motion",
Princeton, University Press, Princeton, 1967

\bibitem{vigier} P. Garbaczewski, J. P. Vigier, Phys. Rev. A 46,
(1992), 4634;

\bibitem{pre99} P. Garbaczewski, "Perturbations of noise: Origins
of isothermal  flows", Phys. Rev. E, in press

\bibitem{blanch}  Ph. Blanchard, P. Garbaczewski, Phys. Rev. E 49,
(1994), 3815

\bibitem{zambrini} J. C. Zambrini, J. Math. Phys. 27, (1986), 2307


\bibitem{geilikman}   B. T. Geilikman, Sov. JETP (Russian edition),
17, (1947), 830

\bibitem{skor} G. A. Skorobogatov, Rus. J. Phys. Chem. 61,
(1987), 509

\bibitem{guth} E. Guth, Adv. Chem. Phys. 15, (1969), 363



\bibitem{carlen} E. Carlen, Commun. Math. Phys. 94, (1984), 293

\bibitem{hol} P. R. Holland, "Quantum theory of motion",
Cambridge University Press, Cambridge, 1993




\end{thebibliography}
\end{document}